\documentclass{desyproc}

\newcommand{\bea}{\begin{eqnarray}}
\newcommand{\eea}{\end{eqnarray}}

\newcommand{\Li}{\mathop{\mathrm{Li}}\nolimits}

\begin{document}
\title{Heavy-quark contributions to the ratio  $F_L/F_2$
at low values of the Bjorken variable $x$ }

\author{{\slshape A.Yu. Illarionov$^1$, B.A.~Kniehl$^2$,
A.V. Kotikov$^3$}\\[1ex]
$^1$SISSA, via Beirut, 2-4, 34014 Trieste and INFN,
Sezione di Trieste, Trieste, Italy\\
$^2$II. Institut f\"ur Theoretische Physik, Universit\"at Hamburg,
22761 Hamburg, Germany\\
$^3$BLThPh, JINR, 141980 Dubna (Moscow resion), Russia}

\contribID{smith\_joe}


\maketitle

\begin{abstract}
We study the heavy-quark contributions to the proton structure functions
$F_2^i(x,Q^2)$ and $F_L^i(x,Q^2)$, with $i=c,b$, for small values of Bjorken's
$x$ variable
and provide compact formulas for their
ratios $R_i=F_L^i/F_2^i$ that are useful to extract $F_2^i(x,Q^2)$ from
measurements of the doubly differential cross section of inclusive
deep-inelastic scattering at DESY HERA.
Our approach naturally explains why $R_i$ is approximately independent of $x$
and the details of the parton distribution functions in the low-$x$ regime.

\end{abstract}

\section{Introduction}

The totally inclusive cross section of deep-inelastic lepton-proton
scattering (DIS) depends on the square $s$ of the centre-of-mass energy,
Bjorken's variable $x=Q^2/(2pq)$, and the inelasticity variable $y=Q^2/(xs)$,
where $p$ and $q$ are the four-momenta of the proton and the virtual photon,
respectively, and $Q^2=-q^2>0$.
The doubly differential cross section is parameterized in terms of the
structure function $F_2$ and the longitudinal structure function $F_L$, as
\begin{equation}
\frac{d^2 \sigma}{dx\,dy}=\frac{2\pi\alpha^2}{xQ^4}
\{[1+(1-y)^2]F_2(x,Q^2)-y^2 F_L(x,Q^2)\},
\label{in}
\end{equation}
where $\alpha$ is Sommerfeld's fine-structure constant.
At small values of $x$, $F_L$ becomes non-negligible and its contribution
should be properly taken into account when the $F_2$ is extracted from the
measured cross section.
The same is true also for the contributions $F_2^i$ and $F_L^i$ of $F_2$ and
$F_L$ due to the heavy quarks $i=c,b$.

Recently, the H1 \cite{Adloff:1996xq,Aktas:2004az,Aktas:2005iw} and ZEUS
\cite{Breitweg:1997mj,Chekanov:2003rb,Chekanov:2007ch}
Collaborations at HERA presented new data on $F_2^c$ and $F_2^b$.
At small $x$ values, of order $10^{-4}$, $F_2^c$ was found to be around
$25\%$ of $F_2$, which is considerably larger than what was observed by the
European Muon Collaboration (EMC) at CERN \cite{Aubert:1982tt} at larger $x$
values, where it was only around $1\%$ of $F_2$.
Extensive theoretical analyses in recent years have generally served to
establish that the $F_2^c$ data can be described through the perturbative
generation of charm within QCD (see, for example, the review in
Ref.~\cite{Cooper-Sarkar:1997jk} and references cited therein).

In the framework of Dokshitzer-Gribov-Lipatov-Altarelli-Parisi (DGLAP)
dynamics \cite{Gribov:1972ri}, there are two basic methods to study
heavy-flavour physics.
One of them \cite{Kniehl:1996we} is based on the massless evolution of parton
distribution functions (PDF) and
the other one on the photon-gluon fusion (PGF) process
\cite{Frixione:1994dv}.
There are also some interpolating schemes (see
Ref.~\cite{Olness:1987ep} and references cited therein).
The present HERA data on $F_2^c$
\cite{Aktas:2004az,Aktas:2005iw,Chekanov:2003rb,Chekanov:2007ch} are in good
agreement with the modern theoretical predictions.

In earlier HERA analyses \cite{Adloff:1996xq,Breitweg:1997mj}, $F_L^c$ and
$F_L^b$ were taken to be zero for simplicity.
Four years ago, the situation changed:
in the
papers
\cite{Aktas:2004az,Aktas:2005iw,Chekanov:2003rb,Chekanov:2007ch},
the $F_L^c$ contribution at
next-to-leading order (NLO) was subtracted from the data.

In this paper,
we present compact low-$x$ approximation formulae \cite{Illarionov:2008be}
for the
ratio $R_i=F_L^i/F_2^i$ at leading order (LO) and NLO, which greatly simplify
the extraction of $F_2^i$ from measurements of
$d^2 \sigma^{i\overline{i}}/(dx\,dy)$.

\section{Parton distribution functions at small $x$}

The standard program to study the small $x$ behavior of
quarks and gluons
is carried out by comparison of the data
with the numerical solution of the
DGLAP
equations
fitting the parameters of the
$x$ profile of partons at some initial $Q_0^2$ and
the QCD energy scale $\Lambda$ (see, for instance, \cite{fits,KKPS}).
However,
in analyzing
exclusively the
small $x$ region ($x \leq 0.01$),
there is the alternative of doing a simpler analysis
by using some of the existing analytical solutions of DGLAP
in the small $x$ limit (see \cite{Kot07} for review).
It was done
in Refs.
\cite{BF1}-\cite{Illarionov:2004nw},
where it was pointed out that the HERA small $x$ data can be
interpreted in
the so called doubled asymptotic scaling (DAS) approximation
related to the asymptotic
behavior of the DGLAP evolution
discovered  in \cite{Rujula}
many years ago.

Here we illustrate results obtained
in \cite{Kotikov:1998qt,Illarionov:2004nw}:
the small $x$ asymptotic PDF
form
in the framework of the DGLAP equation starting at some $Q^2_0$ with
the flat function:
 \begin{eqnarray}
xf_a (x,Q^2_0) ~=~
A_a ~~~~(\mbox{hereafter } a=q,g), \label{1}
 \end{eqnarray}
where $xf_a$ are the leading-twist PDF parts
and $A_a$ are unknown parameters that have to be determined from data.
We neglect
the non-singlet quark component at small $x$.

We would like to note that HERA data \cite{H1} show a rise
of $F_2$
at low $Q^2$ values ($Q^2 < 1 $GeV$^2$)
when $x \to 0$.
This rise can be explained naturally
by incorporation  of higher-twist terms in the
analysis (see \cite{Illarionov:2004nw} and  Fig.1).

\begin{figure}[t]
\vskip -0.5cm
\epsfig{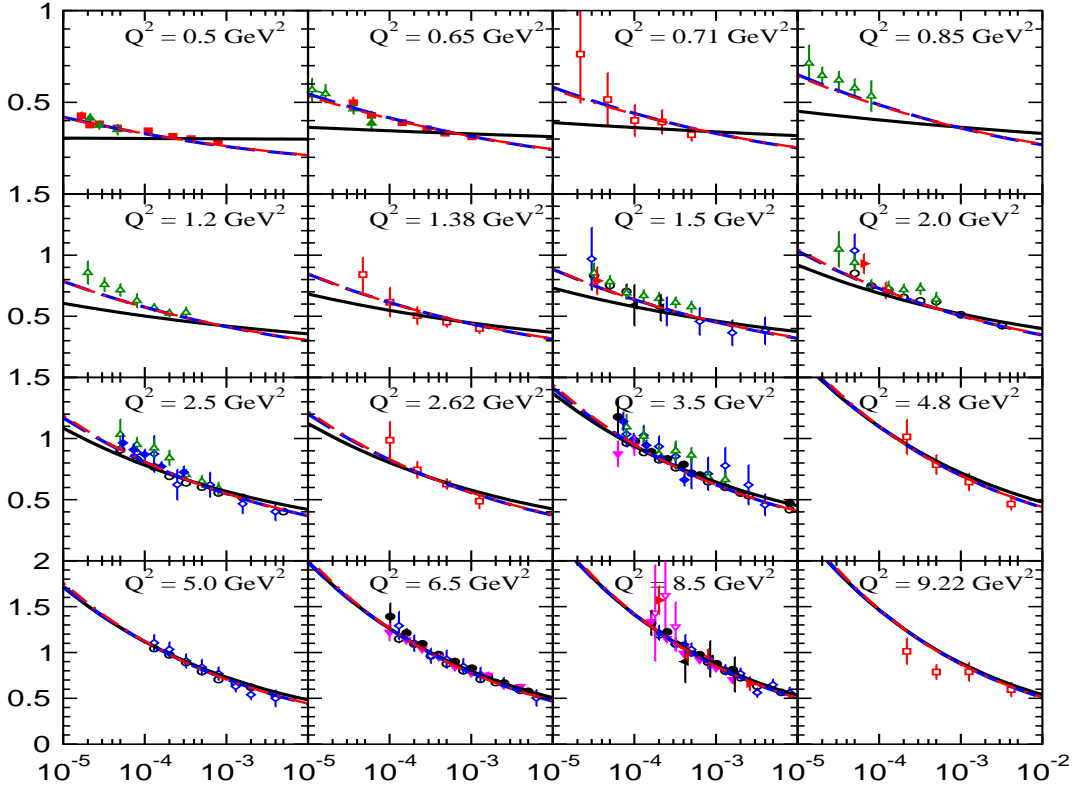}
%
\vskip 1.2cm
\caption{The structure function $F_2$ as a function of $x$ for different
$Q^2$ bins.
The solid and dashed lines are obtained without and with higher-twist
terms, respectively.
}
\end{figure}

We shortly compile
the LO results
(the NLO ones
may be found
in \cite{Kotikov:1998qt,Illarionov:2004nw}), which
are:
 \begin{eqnarray}
 f_a(x,Q^2) &=& f^{+}_a(x,Q^2) +
f^{-}_a(x,Q^2) \; ,
\label{r11}\\
f^{+}_g(x,Q^2)&=& \biggl(A_g + \frac{4}{9} A_q \biggl)
I_0(\sigma) \; e^{-\overline d_{+}(1) s} ~+~O(\rho)
~~\;\; ,\label{8.0} \\
f^{+}_q(x,Q^2)&=& \frac{f}{9}\biggl(A_g + \frac{4}{9} A_q \biggl)
\rho \;
I_1(\sigma) \;
e^{-\overline d_{+}(1) s} ~+~O(\rho) \; , \label{8.01}
\\
f^{-}_g(x,Q^2)&=& - \frac{4}{9} A_q e^{- d_{-}(1) s}
~+~O(x) ,
\label{8.00} \\
f^{-}_q(x,Q^2)&=&  A_q e^{- d_{-}(1) s} ~+~O(x) \; ,\label{8.02}
 \end{eqnarray}
where where $e=(\sum_1^f e_i^2)/f$ is the average charge square and
$\overline d_{+}(1) = 1+20f/(27\beta_0)$ and
$          d_{-}(1) = 16f/(27\beta_0)$
are the regular parts of $d_{+}$ and $d_{-}$
anomalous dimensions, respectively, in the limit $n\to1$
\footnote{
For a quantity $k(n)$ we use the notation
$\hat k(n)$ for the singular part when $n\to1$ and
$\overline k(n)$ for the corresponding regular part. }.
%
%
The functions $
I_{\nu}$ ($\nu=0,1$)
are
the modified Bessel
functions $I_{\nu}$
and the variables $\sigma$ and $\rho$ are
given by
\begin{eqnarray}
\sigma =2\sqrt{\hat d_{+} s \ln(x)} \; , ~~~
\rho = \sqrt{\frac{\hat d_{+} s}{\ln(x)}}
= \frac{\sigma}{2\ln(1/x)}, ~~~ \hat d_{+} = - \frac{12}{\beta_0},
\label{slo}
\end{eqnarray}
where $\beta_0$ is the first coefficient of the QCD beta function and
$s=\ln[a_s(Q_0)/a_s(Q)]$, with $Q_0$ being the initial scale of the DGLAP
evolution, and $a_s(\mu)=\alpha_s(\mu)/(4\pi)$ is the couplant with
the renormalization scale $\mu$.

\section{Master formula}
\label{sec:approach}

We now derive our master formula for $R_i(x,Q^2)$ appropriate for small values
of $x$, which has the advantage of being independent of the PDFs
$f_a(x,Q^2)$.
In the low-$x$ range, where only the gluon and quark-singlet contributions
matter, while the non-singlet contributions are negligibly small, we
have\footnote{%
Here and in the following, we suppress the variables $\mu$ and $m_i$ in the
argument lists of the structure and coefficient functions for the ease of
notation.
Moreover, a
further simplification is obtained by neglecting the contributions due to
incoming light quarks and antiquarks in Eq.~(\ref{eq:pm}), which is justified
because they vanish at LO and are numerically suppressed at NLO for small
values of $x$.
One is thus left with the PGF contribution.
}
\begin{equation}
F_k^i(x,Q^2)=
\sum_{l=+,-}
C_{k,g}^l(x,Q^2)\otimes xf_g^l(x,Q^2),
\label{eq:pm}
\end{equation}
where $l=\pm$ labels the usual $+$ and $-$ linear combinations of the gluon
contributions, $C_{k,g}^l(x,Q^2)$ are the DIS coefficient
functions, which can be calculated perturbatively in the parton model of QCD,
and the symbol $\otimes$ denotes convolution according to
the usual prescription, $f(x)\otimes g(x)=\int_x^1(dy/y)f(y)g(x/y)$.
Massive kinematics requires that $C_{k,g}^l=0$ for $x>b_i=1/(1+4a_i)$, where
$a_i=m_i^2/Q^2$.
We take $m_i$ to be the solution of $\overline{m}_i(m_i)=m_i$, where
$\overline{m}_i(\mu)$ is defined in the modified minimal-subtraction
($\overline{\mathrm{MS}}$) scheme.

Exploiting the low-$x$ asymptotic behaviour of $f_a^l(x,Q^2)$
\cite{Abramowicz:1991xz},
\begin{equation}
f_a^l(x,Q^2)\stackrel{x\to0}{\to}\frac{1}{x^{1+\delta_l}} \tilde{f}_a^l(x,Q^2),
\end{equation}
where the rise of $\tilde{f}_a^l(x,Q^2)$ as $x\to0$ is less than any power of
$x$, Eq.~(\ref{eq:pm}) can be rewritten as
\cite{Lopez:1979bb,Kotikov:1993xe}
\begin{equation}
F_k^i(x,Q^2)\approx
\sum_{l=+,-}
M_{k,g}^l(1+\delta_l,Q^2)xf_g^l(x,Q^2),
\label{eq:pm1}
\end{equation}
where
\begin{equation}
M_{k,a}^l(n,Q^2)=\int_0^{b_i}dx\,x^{n-2}C_{k,a}^l(x,Q^2)
\label{eq:mel}
\end{equation}
is the Mellin transform, which is to be analytically continued from integer
values $n$ to real values $1+\delta_l$ \cite{Kazakov:1987jk}.

In the DAS
approach
\footnote{The singular PDF behavior has been considered recently in
\cite{Ivanov:2008er}.},
one has $M_{k,a}^+(1,Q^2)=M_{k,a}^-(1,Q^2)$ if
$M_{k,a}^l(n,Q^2)$ are devoid of singularities in the limit $\delta_l\to0$, as
we assume for the time being.
Such singularities actually occur at NLO, leading to modifications to be
discussed in Section~\ref{sec:nlo}.
Defining $M_{k,a}(1,Q^2)=M_{k,a}^\pm(1,Q^2)$ and using (\ref{eq:pm}),
Eq.~(\ref{eq:pm1}) may be simplified to
become
\begin{equation}
F_k^i(x,Q^2)\approx M_{k,g}(1,Q^2) \, xf_g(x,Q^2).
\label{eq:pm3}
\end{equation}
In fact, the non-perturbative input $f_g(x,Q^2)$ does cancels in the ratio
\begin{equation}
R_i(x,Q^2)\approx\frac{M_{L,g}(1,Q^2)}{M_{2,g}(1,Q^2)},
\label{eq:ri}
\end{equation}
which is very useful for practical applications.
Through NLO, $M_{k,g}(1,Q^2)$ exhibits the structure
\begin{eqnarray}
M_{k,g}(1,Q^2)~=~
e_i^2 a_s(\mu)\left\{M_{k,g}^{(0)}(1,a_i)
+a_s(\mu)\left[M_{k,g}^{(1)}(1,a_i)+M_{k,g}^{(2)}(1,a_i)
\ln\frac{\mu^2}{m_i^2}
\right]\right\}+{\mathcal O}(a_s^3).
\label{eq:exp}
\end{eqnarray}
where
Inserting Eq.~(\ref{eq:exp}) into Eq.~(\ref{eq:ri}), we arrive at our master
formula
\begin{eqnarray}
R_i(x,Q^2)~\approx ~
\frac{M_{L,g}^{(0)}(1,a_i)+a_s(\mu)
\left[M_{L,g}^{(1)}(1,a_i)+M_{L,g}^{(2)}(1,a_i)\ln(\mu^2/m_i^2)\right]}
{M_{2,g}^{(0)}(1,a_i)+a_s(\mu)
\left[M_{2,g}^{(1)}(1,a_i)+M_{2,g}^{(2)}(1,a_i)\ln(\mu^2/m_i^2)\right]}
+{\mathcal O}(a_s^2).
\label{eq:master}
\end{eqnarray}
We observe that the right-hand side of Eq.~(\ref{eq:master}) is approximately
independent of $x$, a remarkable feature that is automatically exposed by our
procedure.
In the next two sections, we present compact analytic results for the LO
($j=0$) and NLO ($j=1,2$) coefficients $M_{k,g}^{(j)}(1,a_i)$, respectively.

\section{LO results}
\label{sec:lo}

The LO coefficient functions of PGF can be obtained from the QED case
\cite{Baier:1966bf} by adjusting coupling constants and colour factors, and
they read \cite{Witten:1975bh,Kotikov:2001ct}
\begin{eqnarray}
C_{2,g}^{(0)}(x,a) &=& -2x\{[1-4x(2-a)(1-x)]\beta
-[1-2x(1-2a)
+2x^2(1-6a-4a^2)]L(\beta)\},
\nonumber \\
C_{L,g}^{(0)}(x,a) &=&  8 x^2[(1-x)\beta-2ax L(\beta)],
\end{eqnarray}
where
\begin{equation}
\beta(x)=\sqrt{1-\frac{4ax}{1-x}},\qquad
L(\beta)=\ln\frac{1+\beta}{1-\beta}.
\end{equation}

Performing the Mellin transformation in Eq.~(\ref{eq:mel}), we find
(see details in \cite{Illarionov:2008be})
\begin{eqnarray}
M_{2,g}^{(0)}(1,a) &=& \frac{2}{3}[1+2(1-a)J(a)],~~
M_{L,g}^{(0)}(1,a) = \frac{4}{3}b[1+6a-4a(1+3a)J(a)],
\end{eqnarray}
where
\begin{equation}
J(a) = - \sqrt{b}\ln t,\qquad t=\frac{1-\sqrt{b}}{1+\sqrt{b}},
\label{eq:ja}
\end{equation}
At LO, the low-$x$ approximation formula thus reads
\begin{equation}
R_i\approx
2b_i\frac{1+6a_i-4a_i(1+3a_i)J(a_i)}{1+2(1-a_i)J(a_i)}.
\label{eq:lo}
\end{equation}

\section{NLO results}
\label{sec:nlo}

The NLO coefficient functions of PGF are rather lengthy and not published in
print; they are only available as computer codes \cite{Laenen:1992zk}.
For the purpose of this letter, it is sufficient to work in the high-energy
regime, defined by $x\ll1$, where they assume the compact form
\cite{Catani:1992zc}
\begin{equation}
C_{k,g}^{(j)}(x,a)=\beta R_{k,g}^{(j)}(1,a),
\label{eq:nlo}
\end{equation}
with
\begin{eqnarray}
R_{2,g}^{(1)}(1,a)&=&\frac{8}{9}C_A[5+(13-10a)J(a)+6(1-a)I(a)],~~
R_{k,g}^{(2)}(1,a) = -4 C_A M_{k,g}^{(0)}(1,a),
\nonumber\\
R_{L,g}^{(1)}(1,a)&=&-\frac{16}{9}C_A b
\{1-12a-[3+4a(1-6a)]J(a)+12a(1+3a)I(a)\},
\end{eqnarray}
where $C_A=N$ for the colour gauge group SU(N), $J(a)$ is defined by
Eq.~(\ref{eq:ja}), and
\begin{equation}
I(a)=-\sqrt{b}\left[\zeta(2)+\frac{1}{2}\ln^2t-\ln(ab)\ln t+2\Li_2(-t)\right].
\end{equation}
Here, $\zeta(2)=\pi^2/6$ and
$\Li_2(x)=-\int_0^1(dy/y)\ln(1-xy)$ is the dilogarithmic function.

As already mentioned in Section~\ref{sec:approach}, the Mellin transforms of
$C_{k,g}^{(j)}(x,a)$ exhibit singularities in the limit $\delta_l\to0$, which
lead to modifications in our formalism, namely in Eqs.~(\ref{eq:pm3}) and
(\ref{eq:master}).
As was shown in Refs.~\cite{Kotikov:1993xe,Kotikov:1998qt,Illarionov:2004nw},
the terms involving $1/\delta_l$ depend on the exact form of the subasymptotic
low-$x$ behaviour encoded in $\tilde{f}_g^l(x,Q^2)$, as
\begin{equation}
\frac{1}{\delta_l}=\frac{1}{\tilde{f}_g^l(\hat{x},Q^2)}
\int^1_{\hat{x}}\frac{dy}{y}\tilde{f}_g^l(y,Q^2),
\end{equation}
where $\hat{x}=x/b$.
In the generalized DAS
regime, given by Eqs. (\ref{r11})-(\ref{8.02}),
we have
\begin{equation}
\frac{1}{\delta_+}
\approx \frac{1}{\hat{\rho}}\,
\frac{I_1(\sigma(\hat{x}))}{I_0(\sigma(\hat{x}))},
\qquad
\frac{1}{\delta_-}
\approx \ln\frac{1}{\hat{x}}.
\end{equation}
Because the ratio $f_g^-(x,Q^2)/f_g^+(x,Q^2)$ is rather small at the $Q^2$
values considered, Eq.~(\ref{eq:pm3}) is modified to become
\begin{equation}
F_k^i(x,Q^2)\approx\tilde{M}_{k,g}(1,Q^2)xf_g(x,Q^2),
\end{equation}
where $\tilde{M}_{k,g}(1,Q^2)$ is obtained from $M_{k,g}(n,Q^2)$ by taking the
limit $n\to 1$ and replacing $1/(n-1)\to1/\delta_+$.
Consequently, one needs to substitute
\begin{equation}
M_{k,g}^{(j)}(1,a)\to\tilde{M}_{k,g}^{(j)}(1,a)\quad(j=1,2)
\end{equation}
in the NLO part of Eq.~(\ref{eq:master}).
Using the identity
\begin{equation}
\frac{1}{I_0(\sigma(\hat{x}))}
\int^1_{\hat{x}}\frac{dy}{y}\beta\left(\frac{x}{y}\right) I_0(\sigma(y))
\approx \frac{1}{\delta_+}-\ln (ab)-\frac{J(a)}{b},
\end{equation}
we find the Mellin transform~(\ref{eq:mel}) of Eq.~(\ref{eq:nlo}) to be
\begin{equation}
\tilde{M}_{k,g}^{(j)}(1,a)\approx
\left[\frac{1}{\delta_+}-\ln(ab)-\frac{J(a)}{b}\right]
R_{k,g}^{(j)}(1,a)\quad(j=1,2).
\end{equation}
The rise of the NLO terms as $x\to0$ is in agreement with earlier
investigations \cite{Nason:1987xz}.

\section{Results}
\label{sec:results}

As for our input parameters, we choose \cite{Illarionov:2008be}
$Q_0^2=0.306$~GeV$^2$, $m_c=1.25$~GeV and $m_b=4.7$~GeV.
While the LO result for $R_i$ in Eq.~(\ref{eq:lo}) is independent of the
unphysical mass scale $\mu$, the NLO formula~(\ref{eq:master}) does depend on
it, due to an incomplete compensation of the $\mu$ dependence of $a_s(\mu)$ by
the terms proportional to $\ln(\mu^2/Q^2)$, the residual $\mu$ dependence
being formally beyond NLO.
In order to estimate the theoretical uncertainty resulting from this,
in \cite{Illarionov:2008be} we put
$\mu^2=\xi Q^2$ and vary $\xi$.
Besides our default choice $\xi=1+4a_i$, we also considered the extreme choice
$\xi=100$, which is motivated by the observation that NLO corrections are
usually large and negative at small $x$ values \cite{Salam:1998tj}.
A large $\xi$ value is also advocated in Ref.~\cite{Dokshitzer:1993pf}, where
the choice $\xi=1/x^{\Delta}$, with $0.5<\Delta<1$, is proposed.

We now extract $F_2^i(x,Q^2)$ ($i=c,b$) from the H1 measurements of the
cross sections in Eq.~(\ref{in}) at low ($12<Q^2<60$~GeV$^2$)
\cite{Aktas:2005iw} and high ($Q^2>150$~GeV$^2$) \cite{Aktas:2004az} values of
$Q^2$ using the LO and NLO results for $R_i$ derived in Sections~\ref{sec:lo}
and \ref{sec:nlo}, respectively.
Our NLO results for $\mu^2=\xi Q^2$ with $\xi=1+4a_i$ are presented for
$i=c,b$ in Table~\ref{tab:c},
where they are
compared with the values determined by H1.
We refrain from showing our results for other popular choices, such as
$\mu^2=4m_i^2,Q^2$ and even $\mu^2= 100 Q^2$ because they are very similar.
We observe that the theoretical uncertainty related to the freedom in the
choice of $\mu$ is negligibly small and find good agreement with the results
obtained by the H1 Collaboration using a more accurate, but rather cumbersome
procedure \cite{Aktas:2004az,Aktas:2005iw}.

\begin{figure}[t]
\begin{center}
\epsfig{figure=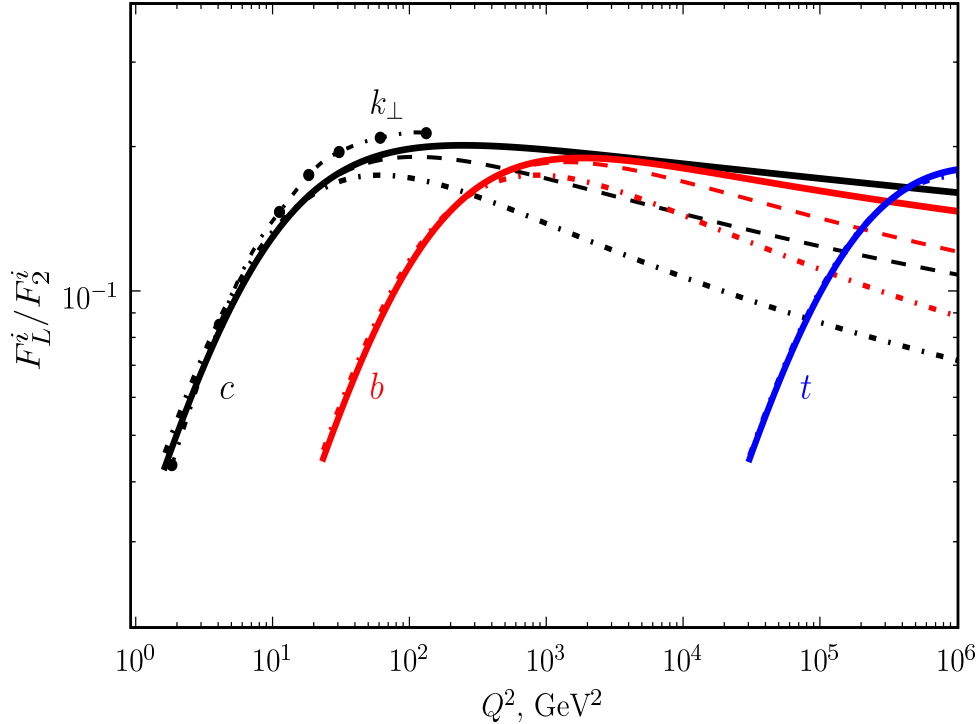,height=4.1in,width=5.6in}
\end{center}
\vskip -1.0cm
\caption{$R_c$, $R_b$, and $R_t$ evaluated as functions of $Q^2$ at LO from
Eq.~(\ref{eq:lo}) (dot-dashed lines) and at NLO from Eq.~(\ref{eq:master})
with $\mu^2=4m_i^2$ (dashed lines) and $\mu^2=Q^2+4m_i^2$ (solid lines).
For comparison, the prediction for $R_c$ in the $k_t$-factorization approach
(dot-dot-dashed line) \cite{Kotikov:2001ct} is also shown.}
\label{fig:r}
\end{figure}

In order to assess the significance of and the theoretical uncertainty in the
NLO corrections to $R_i$, we show in Fig.~\ref{fig:r} the $Q^2$
dependences of $R_c$, $R_b$, and $R_t$ evaluated at LO from Eq.~(\ref{eq:lo})
and at NLO from Eq.~(\ref{eq:master}) with $\mu^2=4m_i^2,Q^2+4m_i^2$.
We observe from  Fig.~\ref{fig:r} that the NLO predictions are rather stable
under scale variations and practically coincide with the LO ones in the lower
$Q^2$ regime.
On the other hand, for $Q^2\gg4m_i^2$, the NLO predictions overshoot the LO
ones and exhibit a strong scale dependence.
We encounter the notion that the fixed-flavour-number scheme used here for
convenience is bound to break down in the large-$Q^2$ regime due to unresummed
large logarithms of the form $\ln(Q^2/m_i^2)$.
In our case, such logarithms do appear linearly at LO and quadratically at
NLO.
In the standard massless factorization, such terms are responsible for the
$Q^2$ evolution of the PDFs and do not contribute to the coefficient
functions.
In fact, in the variable-flavour-number scheme, they are
$\overline{\mathrm{MS}}$-subtracted from the coefficient functions and
absorbed into the $Q^2$ evolution of the PDFs.
Thereafter, the asymptotic large-$Q^2$ dependences of $R_i$ at NLO should be
proportional to $\alpha_s(Q^2)$ and thus decreasing.
This is familiar from the Callan-Gross ratio $R=F_L/(F_2-F_L)$, as may be seen
from its $(x,Q^2)$ parameterizations in Ref.~\cite{GonzalezArroyo:1980wf}.
Fortunately, this large-$Q^2$ problem does not affect our results in
Table~\ref{tab:c}
because the bulk of the H1 data is located
in the range of moderate $Q^2$ values.

\begin{table}
\begin{tabular}{|cc|cc|cc|}
\hline
$Q^2$ & $x$ & $F_2^c(x,Q^2)\cdot 10^3$ (H1) & $F_2^c(x,Q^2)\cdot 10^3$ &
$F_2^b(x,Q^2)\cdot 10^4$ (H1) & $F_2^b(x,Q^2)\cdot 10^4$ \\
\hline
12 & 0.197 & $435\pm78$ & 431 & $45\pm27$ & 45 \\
12 & 0.800 & $186\pm24$ & 185 &  $48\pm22$ & 48\\
25 & 0.500 & $331\pm43$ & 329 & $123\pm38$ & 123 \\
25 & 2.000 & $212\pm21$ & 212 & $61\pm24$ & 61 \\
60 & 2.000 & $369\pm40$ & 368 & $190\pm55$ & 190 \\
60 & 5.000 & $201\pm24$ & 200 &  $130\pm47$ & 130\\
200 & 0.500 & $202\pm46$ & 202 & $413\pm128$ & 400 \\
200 & 1.300 & $131\pm32$ & 130 & $214\pm79$ & 212 \\
650 & 1.300 & $213\pm57$ & 214 & $243\pm124$ & 238 \\
650 & 3.200 & $92\pm28$ & 91 & $125\pm55$ & 125 \\
\hline
\end{tabular}
\caption{\label{tab:c}Values of $F_2^c(x,Q^2)$ and $F_2^b(x,Q^2)$ extracted
from the H1
measurements of $\tilde{\sigma}^{c\overline{c}}$ and
$\tilde{\sigma}^{b\overline{b}}$ at low \cite{Aktas:2005iw}
and high \cite{Aktas:2004az} values of $Q^2$ (in GeV$^2$) at various values of
$x$ (in units of $10^{-3}$) using our approach at NLO for $\mu^2=\xi Q^2$ with
$\xi=1+4a_c$.
The LO results agree with the NLO results
within the accuracy of this table.
For comparison, also the results determined in
Refs.~\cite{Aktas:2004az,Aktas:2005iw} are quoted.}
\end{table}

The ratio $R_c$ was previously studied in the framework of the
$k_t$-factorization approach \cite{Kotikov:2001ct} and found to weakly depend
on the choice of unintegrated gluon PDF and to be approximately $x$
independent in the low-$x$ regime (see Fig.~8 in Ref.~\cite{Kotikov:2001ct}).
Both features are inherent in our approach, as may be seen at one glance from
Eq.~(\ref{eq:master}).
The prediction for $R_c$ from Ref.~\cite{Kotikov:2001ct}, which is included in
Fig.~\ref{fig:r} for comparison, agrees well with our results in the lower
$Q^2$ range, which supports the notion that the $k_t$-factorization approach
partially accounts for the higher-order contributions in the low-$x$ regime.

\section{Conclusions}
\label{sec:conclusions}

In this paper, we observed
a compact formula \cite{Illarionov:2008be} for the ratio $R_i=F_L^i/F_2^i$
of the heavy-flavour contributions to the proton structure functions $F_2$ and
$F_L$ valid through NLO at small values of Bjorken's $x$ variable.
We demonstrated the usefulness of this formula by extracting $F_2^c$ and
$F_2^b$ from the doubly differential cross section of DIS recently measured by
the H1 Collaboration \cite{Aktas:2004az,Aktas:2005iw} at HERA.
These results agree with those extracted in
Refs.~\cite{Aktas:2004az,Aktas:2005iw} well within errors.
In the $Q^2$ range probed by the H1 data, NLO predictions agree very well
with the LO ones and are rather stable under scale variations.
Since we worked in the fixed-flavour-number scheme, our results are bound to
break down for $Q^2\gg4m_i^2$, which manifests itself by appreciable QCD
correction factors and scale dependences.
As is well known, this problem is conveniently solved by adopting the
variable-flavour-number scheme, which we leave for future work.
Our approach also simply explains the feeble dependence of $R_i$ on $x$ and the
details of the PDFs in the low-$x$ regime.

  {\it Acknowledgments.}
One of the authors (A.V.K.)
 would like to express his sincerely thanks to the Organizing
 Committee
for the kind invitation.
He was supported in part,
by Heiserberg-Landau program
and by the Russian Foundation for Basic Research (Grant N 08-02-00896-a).

\end{document}